\documentclass[preprint2]{aastex}

\shorttitle{The Catalog of Edge-on Galaxies}
\shortauthors{Bizyaev et al.}

\begin{document}

\title{The Catalog of Edge-on Disk Galaxies from SDSS. Part I: The Catalog and the Structural Parameters of Stellar Disks}

\author{Bizyaev, D. V.\altaffilmark{1,2} , 
Kautsch, S. J.\altaffilmark{3}, 
Mosenkov, A. V.\altaffilmark{4,5}, 
Reshetnikov, V.P. \altaffilmark{5,6}, 
Sotnikova,~N.~Ya.\altaffilmark{5,6}, 
Yablokova, N. V.\altaffilmark{5}, 
and Hillyer, R.W.\altaffilmark{7} }

\altaffiltext{1}{Apache Point Observatory and New Mexico State University, Sunspot, NM, 88349, USA}
\altaffiltext{2}{Sternberg Astronomical Institute, Moscow State University, Moscow, Russia}
\altaffiltext{3}{Nova Southeastern University, Fort Lauderdale, FL, 33314, USA}
\altaffiltext{4}{Central Astronomical Observatory of RAS, Russia}
\altaffiltext{5}{St. Petersburg State University, Russia}
\altaffiltext{6}{Isaac Newton Institute of Chile, St. Petersburg Branch}
\altaffiltext{7}{Christopher Newport University, Newport News, VA, 23606, USA}

\begin{abstract}

We present a catalog of true edge-on disk galaxies automatically 
selected from the Seventh Data Release (DR7) of the Sloan Digital Sky 
Survey. A visual inspection of the $g$, $r$ and $i$ images of about 
15000 galaxies allowed us to split the initial sample of edge-on 
galaxy candidates into 4768 (31.8\% of the initial sample) genuine 
edge-on galaxies, 8350 (55.7\%) non-edge-ons, and 1865 (12.5\%) 
edge-on galaxies not suitable for simple automatic analysis 
because these objects show signs of interaction, warps, or nearby bright stars 
project on it.
We added more candidate galaxies from RFGC, EFIGI, RC3, and 
Galaxy Zoo catalogs found in the SDSS footprints.   
Our final sample consists of 5747 genuine edge-on galaxies. 
We estimate the structural parameters of the stellar disks 
(the stellar disk thickness, radial scale length, and 
central surface brightness) in the galaxies 
by analyzing photometric profiles in each of the g, r, and i images.  
We also perform simplified 
3-D modeling of the light distribution in the stellar disks of edge-on galaxies
from our sample. 
Our large sample is intended to be used for studying scaling relations in the
stellar disks and bulges and for estimating parameters of the thick disks
in different types of galaxies via the image stacking. In this paper we present 
the sample selection procedure and general description of the sample.

\end{abstract}

\keywords{galaxies: structure, galaxies: edge-on}

\section{Introduction}
Edge-on galaxies provide a unique opportunity for studying the vertical structure of galactic components.
Starting from early studies conducted mostly in the optical bands \citep{kormendybruzual78, burstein79, vdKS81a,vdKS81b,kylafisbahcall87} 
using simple photometric profile fitting, the studies of the vertical structure
of galactic components evolved towards complex modeling 
based on the radiation transfer methods  
\citep{xilouris99,yoachim06,bianchi07,baes11,schechtman12,degeyter13} using multiple UV, optical 
and IR data \citep{popescu00,degeyter13}. Most of the structural studies 
employed limited samples of objects using high quality observations. Large surveys conducted during the last decade 
have made available benefits of observing large samples of interesting objects, which helps in statistical studies of the vertical
structure of galactic disks, bulges, and thick disks \citep{zibetti04,bergvall10}. In this paper we describe our approach to selection of 
true edge-on galaxies from objects observed by Sloan Digital Sky Survey \citep[SDSS]{sdssDR8}. We identified about
six thousand genuine edge-on galaxies with inclination angles not more than a few degrees different from perfect edge-on view. 
Our sample allows statistical studies of the vertical structure parameters of galactic components for the largest 
sample known to date.  We also introduce an on-line catalog of processed SDSS images and of corresponding structural parameters, 
which will simplify further studies of edge-on disk galaxies in the optical bands.  
This paper describes our sample selection procedure and our approach to determination of the stellar disk parameters.
The paper is focused mostly on the stellar disk parameters, while bulges will be considered in the next paper.

\section{The Sample of Edge-on Galaxies}

\subsection{Selection of Candidates to the Initial Sample}

The initial sample of candidates to edge-on galaxies was automatically selected from the SDSS DR7 \citep{sdssDR7} 
using its Catalog Access Server query tools. 
The selection criteria are discussed in detail by  \citet{K06,K06b,K09}. This
selection was based on the axial ratio, angular diameter, magnitude and color limits.
Flagged galaxies and objects with extreme magnitude errors we not included.
The SDSS query was taylored to select relatively bright galaxies with apparent Petrosian magnitudes in the g band 
less than 20 mag using the Petrosian flux; galaxies with angular major-axis diameters larger than 30 arcsec based 
upon isoA\_g, the isophotal major axis given in SDSS in g band; and "flat" galaxies with axis ratio greater than 3 in 
the g-band, which is defined by the isophotal axes isoA\_g divided by isoB\_g. 
The objects are also selected in certain (-0.5 $\leq$ g-r $\leq$ 2) and (-0.5 $\leq$ r-i $\leq$ 2) color ranges. The use of the color ranges in reddening-corrected Petrosian magnitudes allows to prevent the inclusion of galaxies with unusual colors caused by AGNs, instrument flaws, or ghost images. All these selection criteria were applied to the Galaxy table (G.) at the SDSS SkyServer using Structured Query Language (sql). 
The sql query follows:

\begin{verbatim}
petroMag_g < 20 
G.isoA_g/G.isoB_g > 3 
-0.5 <  G.dered_g - G.dered_r < 2  
-0.5 <  G.dered_r - G.dered_i < 2  
G.isoA_g > 37.8
\end{verbatim}

The resulting sample consisted of 18277 unique objects. A brief visual inspection of the images was done to get rid of false detections. 
After that, our final sample of the {\it candidates} to edge-on galaxies included 14983 objects.

\subsection{Visual Inspection and the Final Sample of True Edge-on Galaxies \label{vis_inspection}}

The photometrically calibrated SDSS frames with selected candidates in the 
$g$, $r$, and $i$ bands were taken from the SDSS Data Access Server. 
By the time we started working with the images, SDSS Data Release 8 was issued \citep{sdssDR8}, and the images were downloaded from the DR8. 
The images were cleaned of foreground stars. The star candidates were identified in the images as objects with FWHM from 1 to 1.5 arcsec (typical values
during the SDSS imaging campaign, \citet{sdssDR7} ). The stars in the images were replaced by the median values of pixels 
beyond 3 arcsec (about twice the typical FWHM) from their center. Since the selected images from SDSS are far from really crowded fields, this 
method of cleaning from stars did 
not produce strong artifacts. Just a few cases of projected bright stars were caught in the course of the visual inspection and 
such galaxies were removed from the consideration.
The very central regions of the galaxies were excluded from the cleaning procedure.   
Having initial guess about the galactic center coordinates, we fitted ellipses to galactic isophotes 
at the level of signal-to-noise S/N = 2 per pixel (with the image scale of approximately 0.4 arcsec per pixel ). 
Fitting ellipse to the outer galactic isophotes allowed us to adjust the position of the galactic 
center and to determine the size of the "region of interest" that encompasses the whole galaxy (the encompassing ellipse, hereafter). 
The images then were rotated to align the major axis of the encompassing ellipse with 
the X-axis in the new subframes, and then cropped. This allowed us to make a set of images 
with known geometrical parameter centered on the galaxies, which is 
necessary for the further automatic processing (see \S\ref{red_1d}). 
The images were used  for simplification of our visual inspection
and also became a part of our catalog (see \S\ref{catalog}).

As the next step, all objects were visually classified into groups from the standpoint of further availability for automatic processing. 
The galaxies with clearly seen dust layer, or without signs of non-edge-on spiral arms  were classified as true edge-ons. 
As a result, we selected groups of 
genuine edge-on galaxies, non edge-ons, objects that needed manual pre-processing (e.g. because a bright star nearby did not
allow automatic algorithm to determine parameters of the galaxy correctly), and objects that are not suitable for the automatic processing
described below. In other words, we excluded the objects whose galactic midplane cannot be aligned along the major axis 
of a subimage.
The latter group includes significantly warped edge-on galaxies, and interacting galaxies. This group also includes 
galaxies with with very bright projected stars, whose subtraction would modify significant part of the galactic image.
The initial frames with the objects for the manual processing were then inspected visually, the centers of the galaxies
and their orientation parameters were estimated and properly rotated subframes were made. The galaxies then were classified 
in the way described above and added to the main sample.  
The resultant sample was split into 4768 (31.8\%) true edge-on galaxies, 8350 (55.7\%) non-edge-ons, and 1865 (12.5\%) 
of those objects that need a more complex analysis.
We do not consider the latter group of the objects in this paper. 
The non-edge-on galaxies were also excluded from the further consideration.
This paper is focused on the analysis of the structural parameters of bona fide edge-on, 
non warped and non interacting, galaxies. 

The galaxies were classified into obvious morphological types, from Sa to Irr 
using an automatic algorithm, which was described in detail by \citet{K06,K09}. The major goal of this classification is to assign  morphological Hubble types based upon the size of the bulge component since other morphological properties such as the shape of spiral arms 
are obscured at the edge-on view. We use the concentration index and the ellipticity of the objects for making this automated classification.

The concentration index (CI) is widely used as a classification criterion, reflecting a measure of the spheroidal component in galaxies (e.g., \citet{pranger13,strateva01}). We used the CI provided by the SDSS. It is defined as the ratio of the Petrosian radii (petrorad) that contain 90 and 50 \% of the petrosian flux in the r band \citep{stoughton02}. The CI in SDSS is measured using circular apertures. This leads to a significant flaw in classification of galaxies with a wide range of viewing angles in surveys that observe all types of galaxies with different inclinations ranging from face-on to edge-on view. In our work we focus on purely edge-on disk galaxies. This, in turn, means that all our objects are affected in the same way, and we do not have to deal with normalizing inclination effects to the CI since our sample is carefully selected to consist of only edge-on galaxies. \citet{K06} found that the CI clearly separates galaxies with an apparent bulge from galaxies without a clear bulge component. 

The CI separation values are chosen by visual classification and then have been applied to serve as the limiting values for automated classification. However, pure simple disk galaxies were not detected in a satisfying way so that \citet{K06} introduced a second measure which allowed to select bulgeless disks without any central spheroidal component. This parameter reflects the ellipticity (e) of the galaxies, and was based on luminosity weighed elliptical isophotes.  Also in this case \citet{K06} used visual classification to find the best limiting values in order to distinguish the morphological classes according to the eye inspection. The separation of the morphological types is necessarily somewhat arbitrary, and this is in the nature of the classification itself. 
\citet{K06} choose the limits based on visual classification and applied those to the automatic cataloging, and they also were required to be consistent with similar studies \citep{FGC,RFGC}. Therefore, our classification should act as an indicator of the dominance of the bulge component translated into the common language of Hubble types. Later (in Fig. 8) we will see that the classification reflects the Bulge/Total ratios derived from the 1-D profile fitting, which confirms the classification method described above.

In Table~1 and Table~2 we provide the limiting values of CI and ellipticity, as well as their mean values. Note that
the catalog contains a significant number of early-type spirals because we did not limit our selection to flat and bulgeless, 
late-type spirals due to our choice of selection criteria as discussed in section 2.1. A summary of the fraction of different morphological 
types in the sample is shown in Table~3.

\begin{table} 
\begin{center}
\caption{Concentration index classification criteria and observing values. }
\begin{tabular}{lccc}
\tableline\tableline
Type & Lower Limit & Upper Limit& Mean, error  \\
\tableline
Sa         &  2.70 &     --    & 3.191 $\pm$ 0.011 \\
Sab, Sb &  2.70 &     --   & 2.990 $\pm$ 0.005 \\
Sc         &   2.15 &  2.70 & 2.540 $\pm$ 0.005 \\
Scd       &   2.15 &  2.70 & 2.498 $\pm$ 0.007 \\
Sd         &    --    &  2.70 & 2.403 $\pm$ 0.005 \\
Irr          &    --    &  2.15 & 1.894 $\pm$ 0.030 \\
\tableline   
\end{tabular}
\end{center}
\end{table}

%Table2
%\setcounter{table}{0} 
\begin{table} 
\begin{center}
\caption{Ellipticity classification criteria and observing values. }
\begin{tabular}{lccc}
\tableline\tableline
Type & Lower Limit & Upper Limit& Mean, error \\
\tableline
Sa         &     --     & 0.400 & 0.310 $\pm$ 0.004 \\
Sab, Sb & 0.400  &    --     & 0.666 $\pm$ 0.003 \\
Sc          &    --     & 0.766 & 0.713 $\pm$ 0.003 \\
Scd        & 0.766 &  0.816 & 0.794 $\pm$ 0.002 \\
Sd          & 0.816 &    --     & 0.826 $\pm$ 0.003 \\
Irr           &    --     & 0.816 & 0.367 $\pm$ 0.004 \\
\tableline   
\end{tabular}
\end{center}
\end{table}

%%% Table 3.
%\setcounter{table}{0} 
\begin{table} 
\begin{center}
\caption{Edge-on galaxies in our sample by morphological types. }
\begin{tabular}{lr}
\tableline\tableline
Type & Fraction, Percent \\
\tableline
Sa & 7.2\% \\
Sab, Sb  & 32.2\% \\
Sc  &  19.2\% \\
Scd  &  10.8\% \\
Sd  &  28.8\% \\
Irr  &  1.8\% \\
\tableline   
\end{tabular}
\end{center}
\end{table}

Since the initial automatic selection of the galaxy candidates did not include significant number of the largest edge-on galaxies, we 
incorporated more edge-on candidates by supplying objects from the 
Revised Flat Galaxy Catalog \citep[][RFGC, hereafter]{RFGC},  
RC3 \citet{RC3}, EFIGI \citep{EFIGI}, and GalaxyZoo \citep{GalaxyZoo} that were found in the SDSS footprints.
The reason for missing extended objects in the SDSS fields lies in the difficulty to assign correct borders to large and 
extended objects with multiple luminosity centers, e.g., HII regions. This so-called shredding causes that those missed 
galaxies are detected as two or more entries by SDSS's image processing pipeline.

\begin{figure}
%\epsscale{1.0}
\plotone{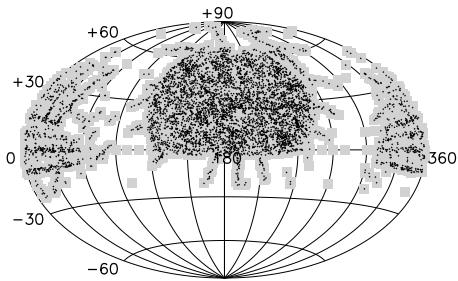}
\caption{The distribution of genuine edge-on galaxies in the sky shows that they cover RA-Dec space essentially the same way as 
all SDSS objects do, according to \citet{sdssDR8} and sdss3.org. The grey shaded area designates the SDSS imaging 
footprints.
\label{fig0}}
\end{figure}

Only the galaxies with major axis greater 
than 30 arcsec (according to the HyperLeda database, \citet{LEDA}) were added.
All new edge-on candidates were processed in the way described above and then visually inspected. 
The final sample of unique true edge-on galaxies consists of 5747 objects. 
Fig.~\ref{fig0} shows that the galaxies in our sample are more or less uniformly allocated in the SDSS imaging survey area.  

\begin{figure}
\epsscale{.70}
\plotone{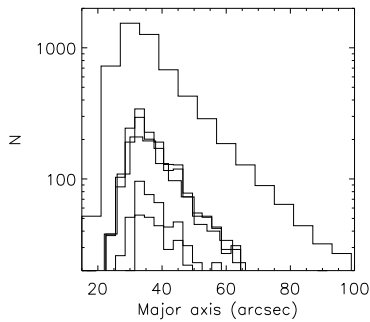} 
\caption{The distribution of the galactic major axis size that was estimated from the SDSS images at S/N=2 level, see text.
 The distribution (top curve) suggests that our sample is complete for the galaxies with the major axis greater than 28 arcsec.
 Different morphological types are shown below the general sample and correspond to Sb, Sd, Sc, Scd, and Sa 
 from top to bottom, respectively.
 \label{fig1}}
 \end{figure}

We checked the completeness of our catalog using the recipes by \citet{thuan_seitzer79}. Their V/V$_m$ value was calculated 
for the whole sample and we found that our genuine edge-on subsample is 95 \% complete for all galaxies with major axis size larger than 
28 arcsec. Fig.~\ref{fig1} shows the histogram of the distribution of the major axes, top curve. Our morphological classification allows for 
splitting the histogram by types, which is shown in Fig.~\ref{fig1}.
Galaxies of different morphological types observed edge-on may be affected by dust on a different way, 
but we see that different types follow the general distribution. We also calculated the Thuan \& Seitzer's V/V$_m$ in dependence of the 
morphological types and found that the Sa, Sb, Sc, Scd, and Sd types are 95\% complete for the galaxies with 
the major axes greater than 33, 28, 26, 25, and 27 arcsec, respectively. The Irr subsample has low statistics not 
enough for certain V/V$_m$ calculation.  Thus, we conclude that the completeness is 
not a strong function of the morphological type.

\begin{figure}
\epsscale{.70}
\plotone{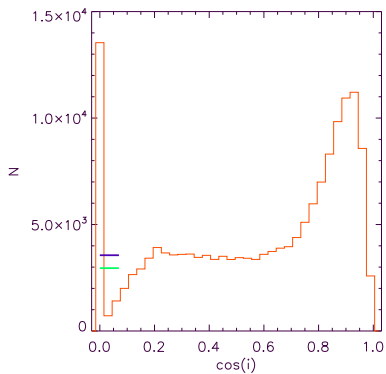}
\caption{Cosine of the inclination angle in the galaxies selected from SDSS DR7 formally estimated from the major-to-minor 
axes ratio (see text) is shown as the solid line histogram. The lower short bar at the highest inclinations designates 
the number of our true 
edge-on galaxies estimated from the objects selected with SDSS CAS query. The upper short bar designates the total 
number of the real edge-on galaxies in our final sample. 
\label{fig2}}
\end{figure}

We performed a test how well we select the true edge-on galaxies among objects with arbitrary inclination. 
For this purpose we submitted a DR7 CAS query 
with the same sample selection criteria except the axis ratio limit. 
As a result, the output comprises 161571 objects. 
The inclination angle $i$ was coarsely estimated from the minor-to-major axes ratio (b/a) using 
equation $\cos^2(i) = ( (b/a)^2 - q^2) / (1- q^2)$ \citep{hubble26}, where we assume that
the intrinsic axial ratio in the galactic disks is the same for all objects and is equal to $q = 0.13$  \citep{Giovanelli94},
and the major and minor axes $a$ and $b$ are the $r$-band sizes from the DR7 CAS tables. 
Fig.~\ref{fig2} shows the distribution of the {\it formally} estimated $\cos(i)$, which has to 
be flat in the case of equal probability of the galactic inclinations in space. 
Fig.~\ref{fig2} demonstrates that the variety of edge-on galaxies cannot be described by a single universal
value of the internal flatness $q$.  
It is seen that due to contamination from non-edge-on galaxies with the inclination between 80 and 86 degrees the inclination distribution 
strongly peaks at 90 degrees. At the same time, the object migration to the peak creates the dip between 80 and 86 degrees. 

Our visual inspection allowed us to select statistically reasonable sample of true edge-on galaxies:
assuming that we classify a galaxy as a genuine edge-on if its inclination is over 86 degrees \citep{degrijs97,BK04},
the corresponding number is calculated from the size of our true-edge-on sample and is 
designated by the lower short bar at 86-90 degrees in Fig. \ref{fig2}. The upper short bar shows the number of the galaxies 
in our full resulting sample (i.e. the sample with further additions beyond the automatic CAS SDSS selection).
The shape of the distribution is the same as in other studies based on SDSS samples, e.g. as in 
\citet{masters_10}. Note that the peak at the left side of the distribution (where cos(i) is close to 1) shows growing
contribution of elliptical galaxies to the general sample. This does not affect the high-inclination
end of the diagram. Note that Fig. \ref{fig2} is the illustration only that shows selection effects in the case
if an oversimplified approach to the inclination determination is applied. The 
procedure of our visual inspection provides a more robust way of selection of the edge-on galaxies.

\section{The Structural Parameters from the 1-D analysis of Photometric Profiles \label{red_1d}}

We performed the analysis of 1-D photometric profiles using the same technique 
as described by \citet{BM02, BM09}. The volume brightness in the stellar disks 
is assumed to change  as follows in the radial $r$ and vertical $z$ 
directions: 
\begin{equation}
\label{vol_disk}
\rho_L(r,z) = \rho_{L_0} \exp(-r/h)\, \mathrm{sech}^2(z/z_0),
\end{equation}
where $h$ is the scalelength and $z_0$ is the scaleheight of the disk. The 
central face-on disk luminosity 
$\displaystyle I_0 = \int_{-\infty}^{\infty} \rho(0,z) dz$ 
corresponds to the central surface brightness $S_0$. The 
model photometric profiles are obtained by the integration of 
equation~(\ref{vol_disk}) along the line-of-sight and by the convolution with the 
instrumental profile. We assume that the point spread function (PSF) is equal 
to 1.4 arcsec for all considered images, given the survey photometry campaign 
description \citep{sdssDR8}. The central regions of the galaxies (1/4 of the 
semimajor axis of the encompassing ellipse) in which bulges can be seen are 
excluded from the stellar disk parameters evaluation in all galaxies. The best-fit scales and 
surface brightness are estimated from the radial or vertical profiles that
were drawn through each pixel row and column within the encompassing ellipses 
in the cleaned and rotated subframes. Although we did not take the the inclination angle
into account in this 1-D approach, our visual inspection should select the galaxies with
small deviation from the 90-degrees inclination. As it has been shown by \citet{barteldrees94,degrijs98,kregel02},
the small deviation from the perfect edge-on view does not affect the structural 
parameters significantly.

We also evaluate and find the structural parameters separately in two halves of the galactic images 
(as seen above and below the galactic midplane) .
An axisymmetric and dustless galaxy, even observed at a small angle with respect to the
perfect edge-on view, will have the same brightness for the parts seen above and below the 
galactic midplane. In the presence of dust extinction the galaxy's halves above and below the
midplane look different for observers, see e.g. \citet{xilouris99,bianchi07}.
To mitigate effects of the dust, we consider the scales and 
surface brightness only for the brightest half of each galaxy. Note that 
because of our selection procedure there is no big difference between the 
parameters determined from the brightest half only and from the 
entire galaxy. The output structural parameters reported by us are the median values of 
all considered photometric profiles.

Once the structural parameter difference in the NIR photometric bands 
is affected by the dust attenuation \citep{BM09}, the difference in the optical bands 
is affected also by the gradients of the stellar population.
This makes us consider the structural parameters 
in the $g$, $r$, and $i$ bands separately from each other.

We also coarsely estimate the contribution of the bulge to the luminosity of 
the galaxies. Using the estimated $S_0$, $h$, and $z_0$, we create images of 
edge-on disks, and subtract them from the images of the galaxies. The 
bulge-to-total luminosity ratio is found as the luminosity of the residual 
image integrated over the region within 1.0 $h$ from the center normalized by 
the total luminosity of the galaxy integrated within the encompassing ellipse.
The structural parameters estimated from the 1-D profile analysis 
for our sample of genuine edge-on galaxies are shown in Table~4.

%% Table 2 should be here.

Comparison with the RFGC shows that our catalog has 917 RFGC objects. 
The vertical and radial sizes of the encompassing ellipses well correlate 
with the size of the galaxies visually estimated in RFGC: our $r$-band 
semimajor axis size is 0.95 $\pm$ 0.13 of the red semimajor axis for the
common galaxies, and our semi minor size is 1.27 $\pm$ 0.21 of the semiminor 
axis for them. We matched our catalog to the sample by \citet{BM09} and found 
53 objects in common. Our $h$ and $z_0$ well correlate with the same structural 
parameters determined by \citet{BM09} for those 53 galaxies: on average, our 
radial scale length in the $i$ is 1.11 $\pm$ 0.43 of that in the $J$ band, and 
the vertical scale height is 1.20 $\pm$ 0.15 of the $J$-band scale from 
\citet{BM09}. A little worse agreement is for 20 common galaxies with
\citet{MSR10}: the ratio of our scale length to the published one is 1.11 $\pm$ 0.25,
and the ratio of the scale heights is 1.42 $\pm$ 0.41 for the J-band images.

Sample by \citet{yoachim06} has 18 common galaxies with our catalog. Our 
structural parameters were determined using the same functional form for the 
radial and vertical profiles, and the results are in reasonable agreement: 
our radial scale length is 1.23 $\pm$ 0.30, and our vertical scale height is 
0.99 $\pm$ 0.13 of those from the R-band estimates by \citet{yoachim06}. 

The parameters for one common with \citet{pohlen_04} thick early type disk 
galaxy are also in a good agreement (1.03 and 1.07 for the radial and vertical 
scales ratio, respectively, between our $g$ and their $V$ images). Our scale 
length is 1.07 $\pm$ 0.36 of that found for the large galaxies by \citet{bianchi07}.
On the other hand, our scale height is much thicker than that reported by \citet{bianchi07}.

\subsection{The Structural Parameters}

\begin{figure}
\epsscale{.90}
\plotone{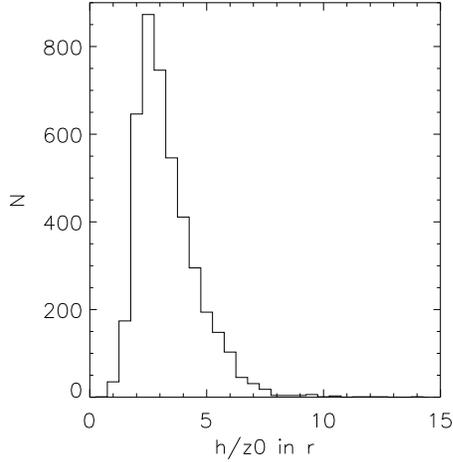}
\caption{The distribution of the inverse stellar disk thickness $h/z_0$ estimated from the 
$r$-band images for all galaxies whose $z_0$ is greater than 3 pixels. 
\label{fig3}}
\end{figure}

The histogram of the distribution of the radial-to-vertical scale ratios 
$h/z_0$ is shown in Fig. \ref{fig3}. The distribution has a prominent peak at 
$h/z_0 \approx 2.5$ in the $r$ band. The median values over the whole sample
are 3.6, 3.4, and 3.3 for the $g$, $r$, and $i$ bands, respectively.
The radial-to-vertical scale ratios in Fig. \ref{fig3} are somewhat 
lower than those for the typical edge-on galaxies estimated from the NIR 
\citep{BM02, BM09}. For the sample of 153 galaxies composed by \cite{BM02} the 
median ratio of $h/z_0$ is about 4.8. This is expected because we 
do not constrain the bulge contribution and consider a wide range of 
disk galaxies (see Fig.~\ref{figBT}), whereas \citet{BM02, BM09} focused 
on bulgeless galaxies. The scale ratios in our sample are consistent with 
the NIR data presented by \cite{MSR10}. 
\cite{MSR10} constructed a sample of 175 edge-on galaxies both of early- and 
late-types and found the median ratio $h/z_0$ to be about 3.5 in the $J$ 
band and 3.9 for the $H$ and $K_\mathrm{s}$ bands.

\begin{figure}
\epsscale{.90}
\plotone{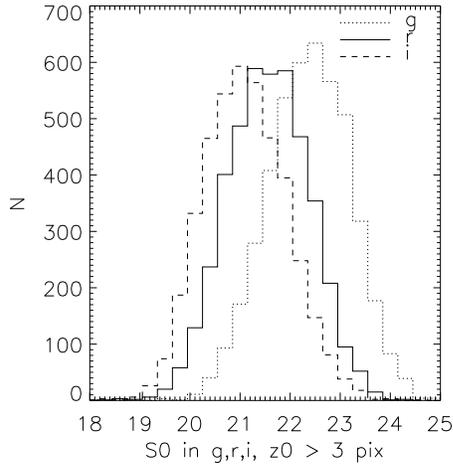}
\caption{The distribution of the stellar disk's central face-on surface brightness. 
The dotted, solid, and dashed curves designate the distributions 
in the $g$, $r$, and $i$ bands, respectively. The surface brightness was corrected for the 
foreground reddening in our Galaxy.
\label{fig4}}
\end{figure}

\begin{figure}
\epsscale{.90}
\plotone{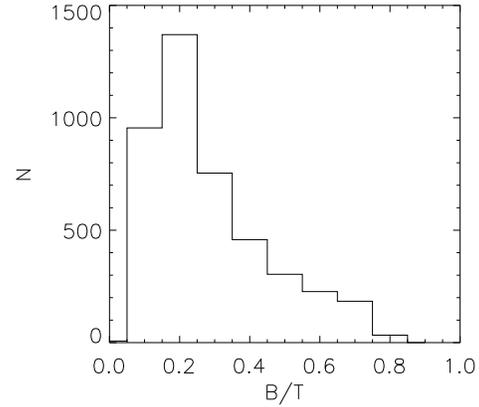}
\caption{The distribution of the bulge-to-total luminosity ratio $B/T$. The ratio is 
estimated from the $r$-band images of all galaxies whose $z_0$ is greater 
than 3 pixels. 
\label{figBT} }
\end{figure}

The distribution of the face-on central surface brightness $S_0$ 
in the $gri$ bands is shown in Fig. \ref{fig4}. The surface brightness 
was corrected for reddening in our Galaxy using the extinction maps by 
\citet{Schlegel_98}. It is seen that the central surface brightness values 
$S_0$ span five magnitudes in each band. 
Apparently, our central surface brightness is biased towards 
the dimmer values with respect to those for arbitrary inclined galaxies,
which is a manifestation of the dust extinction 
within the galaxies. It suggests that a more complex 
modeling will help in better recovery of the stellar disk central brightness
from the data of optical photometry. 

Although the bulge contribution is a parameter affected by dust and 
projection effects in edge-on galaxies, the bulge-to-disk luminosity ratio
is helpful in morphological classification since the spiral arms cannot be observed in edge-on stellar
disks. We show the B/T ratio for the galaxies in our sample in Fig. \ref{figBT} with 
a warning of using this value with caution for direct comparison with arbitrary inclined 
galaxies. A more honest B/T ratio can be recovered from 3-D modeling
with including the central area of galactic images into analysis. 
The inverse scale ratio $h/z_0$ reveals some dependence of the bulge-to-total 
ratio, whereas our 3-D analysis shows that the trend is much less significant.

\begin{figure}
\epsscale{.80}
\plotone{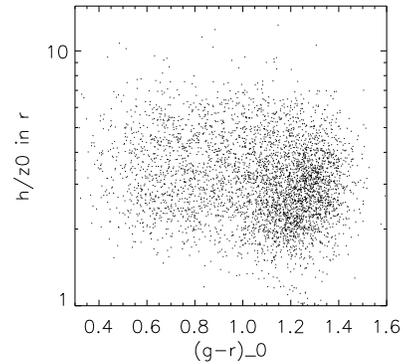}
\caption{The inverse stellar disk thickness $h/z_0$ (in the r-band) 
versus the corrected color (g-r) for all galaxies in our sample. 
The colors are corrected for reddening in our Galaxy.
\label{fig5c}}
\end{figure}

The disk central surface brightness determined for the brighter and dimmer halfs of the 
galaxies allows us to introduce the "brightness asymmetry" parameter estimated as
the difference between the brighter and the dimmer values of the central surface 
brightness. Although it is a function of the galactic inclination in the case of an 
individual galaxy, in a combination with variable disk thickness, 
bulge strength, and clumpy nature of the dust layer it is not a direct measure of the 
inclination in a sample of edge-on galaxies. Nevertheless, the asymmetry gives a possibility to test if our
selection procedure is biased and gives preference to certain values of the asymmetry. 
We ran the V/V$_m$ test as in \S2.2 for several groups of galaxies with similar
asymmetry $Asy$ in the $r$-band: $0 < Asy < 0.14~mag$,   
$0.14~mag < Asy < 0.30~mag$ ,  $0.30~mag < Asy < 0.49~mag$,  
$0.49~mag < Asy < 0.75~mag$,  and $Asy > 0.75~mag$. Each of the five groups
contains approximately equal number of the members, which is 1/5 of the whole sample. 
The 95\% completeness level is achieved for the major axis greater than 30, 29, 29, 28, and 27
arcsec, respectively. It indicates that the galaxies with different asymmetries have  
mostly the same completeness as the whole our sample of galaxies. 
We checked if the asymmetry is a function of 
the disk thickness, central surface brightness, or distance (for those galaxies with available
radial velocities), and did not find any such trends. Thus, we select the objects uniformly 
from the standpoint of their bright - dim halfs asymmetry.

We also sorted out the galaxies by their inverse disk scale ratio $h/z_0$ in the $r$ band and
considered completeness using the V/V$_m$ test. The $h/z_0$ ranges of (0.0, 2.4), (2.4, 3.1,) 
(3.1, 3.9), (3.9, 5.1), and over 5.1 
made the groups of mostly equal size, and their 95\% completeness level starts at the major axis
size of 29, 28, 29, 30, and 28 arcsec, respectively. This shows that our sample is not biased over
the internal disk thickness.

The galactic colors were estimated using equal areas within the encompassing 
ellipses. The colors were corrected for the Milky Way reddening using 
the maps from \citet{Schlegel_98}, but they were not corrected for the internal 
extinction in the galaxies.
We do not observe significant trends in the disk thickness with the overall 
galaxy color, although red objects tend to possess thicker disks, similar 
to that reported by \citet{K09b}.

\begin{figure}
\epsscale{.80}
\plotone{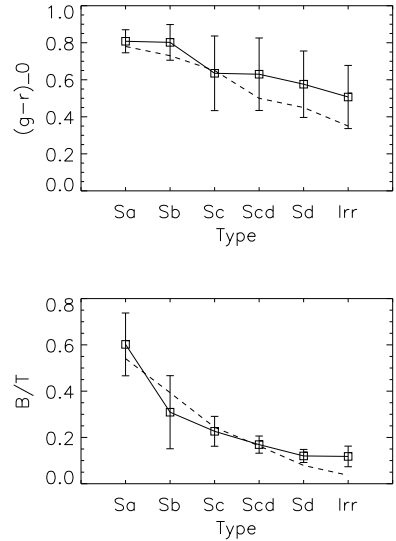}
\caption{The mean values and r.m.s. scatter of the (g-r) color (top) and
of the bulge-to-disk luminosity ratio (bottom) for different morphological types
in our sample. 
The dashed curve in the top panel shows the color morphological type dependence
for a same of isolated galaxies with arbitrary inclination from \citep{AMIGA}.
The dashed curve in the bottom panel denotes the bulge-to-total luminosity
distribution from EFIGI sample \citep{EFIGI}.
\label{color-bt-type}}
\end{figure}

The colors, as well as the bulge-to-disk luminosity ratio, correlate well with the 
estimated morphological type in our sample, see Fig. \ref{color-bt-type}.

More details on the statistical properties of the derived structural parameters 
are presented in Table~4.

%% Table 3 - median values of all parameters in the $gri$ bands.

\subsection{The Scale Height Gradient}

\begin{figure}
\epsscale{.80}
\plotone{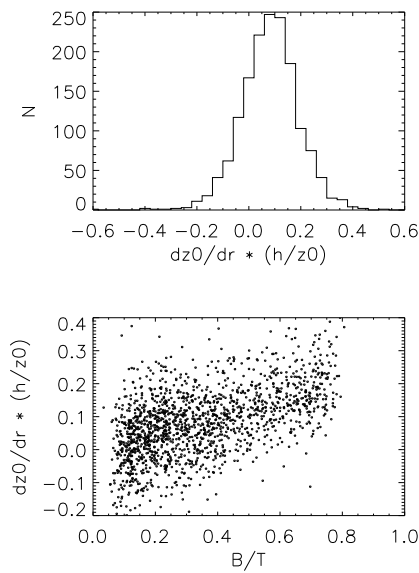}
\caption{Top: The radial gradient of the stellar disk thickness $z_0$ estimated from the r-band images. 
The gradient $dz_0/dr$ is normalized by the disk thickness $z_0/h$. Bottom: the normalized 
radial gradient of the scale height versus the bulge-to-disk ratio. 
\label{fig5}}
\end{figure}

Since we analyzed the vertical photometric profiles independently of each other, we can 
estimate how the vertical scale height changes with the distance to the center 
in terms of the scale height radial gradient. The gradient is calculated from individual vertical 
photometric profiles in the range from 1 to 3 radial scale lengths. 
Fig.~\ref{fig5} (lower panel) shows the distribution of the 
scale height gradient for the whole sample. The gradient was normalized and is shown as
$(d z_0 / dr)/(h/z_0)$. 
The distribution peaks close to zero, at small positive values of the gradient:
the mean value is 0.063, the median is 0.064, and the mode is 0.067.
The fraction of galaxies with strong positive and negative radial gradients of the $z_0$ ($\pm$ 0.2 from the median 
value) is 10.8\%. Interesting that large positive gradients of the scale height is observed 
mostly in the galaxies with significant bulges, whereas bulgeless galaxies have disks 
without the radial gradients of the vertical scale, on average. This trend 
suggests that large bulges affect our method of the gradient determination,
and the gradients inferred for the galaxies with B/T $>$ 0.4 are biased. For the galaxies
with B/T $<$ 0.4 the normalized radial gradient in the r-band is 0.045 on average. 
Given the bulge contamination and dust effects, we should warn the readers about the
limitations of the gradient determination, especially in the case of the smallest galaxies in our sample. 
The vertical scale gradients can be addressed with additions to a 3-D modeling approach, which we will
introduce to the modeling in the next paper. 

\subsection{The Catalog of True Edge-on Galaxies \label{catalog}}

We present our sample of edge-on galaxies as an on-line catalog EGIS (Edge-on Galaxies In SDSS) 
public available at \\ {\it http://users.apo.nmsu.edu/$\sim$dmbiz/EGIS/ }.
The catalog's core table is our Table~4, which contains the structural parameters for each image in the 
$g$, $r$, and $i$ bands. The catalog also contains cleaned images used for our analysis of structure, as well as
raw (not cleaned) images. All images are trimmed to have the galactic center at the center of the frame and 
are rotated to place the major axis parallel to the image rows. 
Note that all initial images for the analysis are taken from the SDSS, so all SDSS
data usage policies are applied to our catalog.  

\begin{figure}
\epsscale{1.1}
\plotone{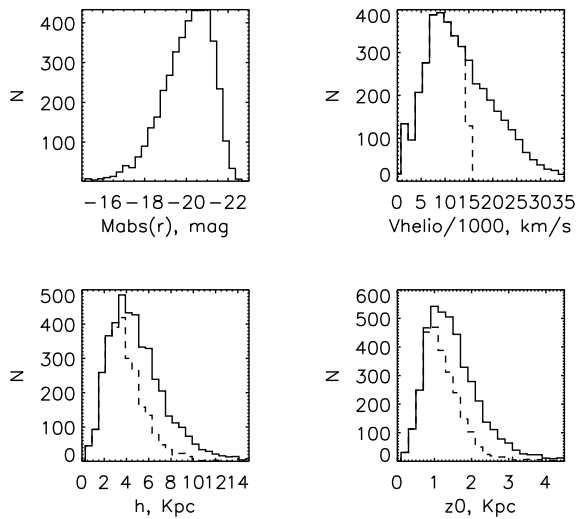}
\caption{The distribution of the 
absolute magnitude in the $r$ band, radial velocity, and the radial ($h$) and vertical ($z_0$) scales expressed in 
kpc. The absolute magnitude is corrected for reddening in the Milky Way.
The $h$ and $z_0$ distributions for the whole sample are designated by the solid line, whereas the dashed line marks 
those distributions for relatively nearby galaxies with heliocentric velocity $V_{helio} < $ 15000 km/s.
\label{fig7_general}}
\end{figure}

Cross-matching over the HyperLeda database allows us to find radial velocities for about 3/4 of our sample. 
The distribution  of the physical properties (the absolute magnitude in the $r$ band, the radial velocity, and the 
radial and vertical scales in the physical units) 
is shown in Fig.~\ref{fig7_general}. 

The future implications of the catalog include co-adding images in order to study properties 
the thick disks statistically, evaluation of the bulge properties, 
and study of scaling relations based on the large sample of similar objects.

\section{Structural Parameters from 3-D Multicomponent Modeling}

Realistic modeling of edge-on galaxies for the structural parameters determination must include the dust extinction. 
As modeling of some large galaxies shows \citep{xilouris99,popescu00,yoachim06,bianchi07}
the dust in the galaxies can be successfully approximated
as embedded disk with uniform extinction and scattering. As \citet{bianchi07} noticed, neglecting the dust scattering 
and taking into account only dust absorption does not 
introduce significant errors to the resulting structural parameters. We simplify calculations and neglect the dust 
scattering in the modeling described below and assume that the dust extinction volume density 
can be expressed as 
\begin{equation}
\label{vol_dust}
\kappa_{\mathrm{ext},\lambda}(r,z) = \frac{\tau_{0,\lambda}}{2 z_\mathrm{d}} 
\exp(-r/h_\mathrm{d})\, \exp(-|z|/z_\mathrm{d}),
\end{equation}

Here $\tau_{0,\lambda}$ is the face-on optical depth of the dust disk at the center, and  $h_\mathrm{d}$ and 
$z_\mathrm{d}$ are the radial and vertical scales of the dust disk. 
The distribution of the luminosity density in our model stellar disks follows equation (\ref{vol_disk}). 
We also add a stellar bulge to the modeling since we did not limit our consideration by bulgeless galaxies and 
many of our galaxies have noticeable bulge (according to Fig. \ref{figBT} and Table~5).
We assume the Hubble volume luminosity density distribution for the 
bulge \citep{Xilouris_98}, which can be written 
\begin{equation}
\label{vol_bulge}
 \rho_\mathrm{b}(r,z) = \rho_\mathrm{0b} (1 - B^2)^{-3/2}~, 
\end{equation}
where $B = \displaystyle \frac{ \sqrt{r^2 + z^2 (b/a)^2} } {R_\mathrm{e}}$, 
$\rho_\mathrm{0b} $ is the central luminosity density in the bulge, and ${R_\mathrm{e}}$ is the bulge effective radius.

After adding the co-planar embedded extinction disk and tilting the system by close to edge-on
inclination angle, we calculate the final brightness distribution via numerical integration of the 
luminosity volume density along the line-of-sight. The 2-D images were convolved with a Gaussian
PSF, same as we used in \S~3.

Attempts of modeling the stellar disk using SDSS images with unconstrained set of parameters was performed by 
\citet{B07} using chi-square minimization of the difference between the real and model galactic images. 
As modeling of small samples of well spatially resolved galaxies shows \citep{xilouris99,matthews99,yoachim06,bianchi07},
the dust disk described by equation (\ref{vol_dust}) has two times smaller scale height than that in the stellar disks. 
To simplify calculations, we assume that $z_\mathrm{d} = z_0/2$ in the further 3-D modeling. 

Our model has 14 free parameters: the X-Y position of the center in the sky plane, 
the position angle of the major axis (PA), the inclination of the galactic plane, central surface 
brightness and the scales of the stellar disk, central face-on extinction and the scales of the 
dust disk, and central surface brightness, axes ratio and effective radius of the bulge.  

\begin{figure}
\epsscale{1.1}
\plotone{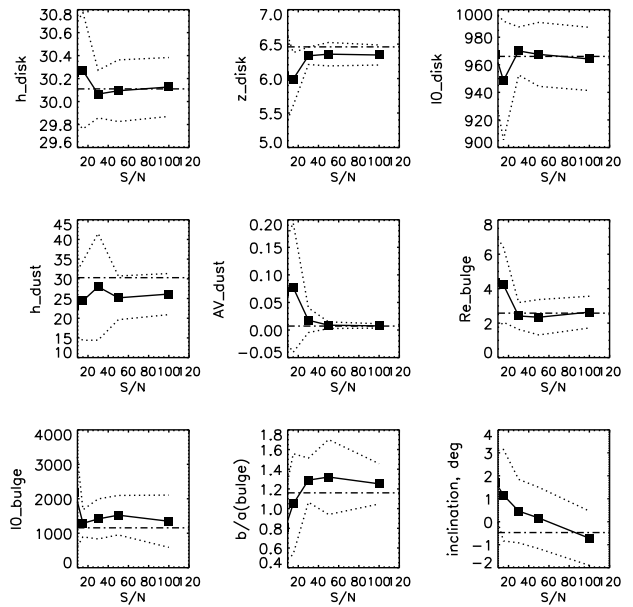}
\caption{The model parameters recovered from our synthetic images. The solid curve with symbols
shows the averaged output parameters, The dashed line designates the input value. The dotted lines designate
the 1-sigma uncertainty of the parameters estimated from 30 simulations for each S/N ratio.  \label{fig9}}
\end{figure}

To ensure that we can reliably recover the galactic component parameters, we performed 
Monte-Carlo simulations and created a set of synthetic images of edge-on galaxies. 
Using equations (\ref{vol_disk}, \ref{vol_dust}, \ref{vol_bulge}), we made a large set of synthetic images and 
projected them to the sky plane by adding some inclination different from the perfect edge-on view and a small position angle tilt. 
A random noise was added to the images in order to degrade the image quality and to make them have 
S/N comparable to the observing data. The synthetic images were then evaluated 
and recovered parameters were compared to the input ones, see Fig. \ref{fig9}.

As the modeling in Fig. \ref{fig9} shows, not all parameters are recovered equally well:
the structural parameters of the stellar disks are the most reliable ones.
 The dust disk
parameters are not reliably estimated. The inclination of the galactic plane is the hardest parameter to estimate even
from a smooth and non clumpy synthetic images. 

\begin{figure}
\epsscale{1.1}
\plotone{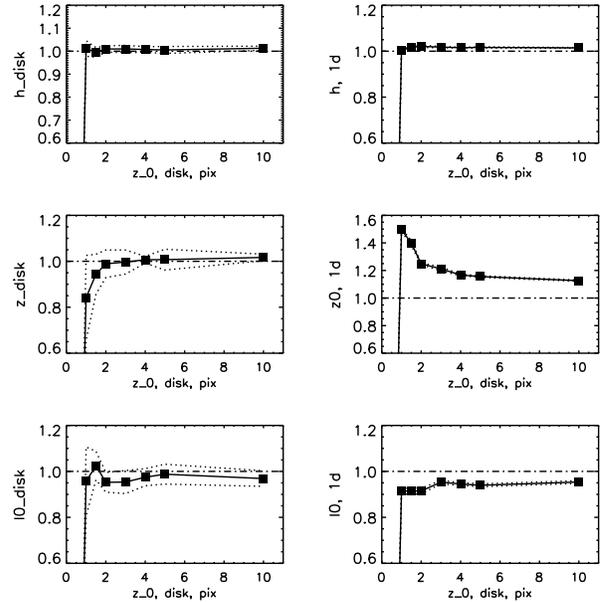}
\caption{The radial and vertical scales and the central surface brightness (in the arbitrary linear units) estimated 
from synthetic images in dependence of the vertical scale height expressed in pixels. The left-side panels
show results of our 3-D analysis, and the right-side panels demonstrate the 1-D analysis. 
The solid curve with symbols
shows the averaged output parameters, The dashed line denotes the input value. The dotted lines designate
the 1-sigma uncertainty of the parameters estimated from 21 simulations for each S/N ratio.  \label{fig9_2}}
\end{figure}

We ran the same simulations with variable pixel size of the synthetic images to understand how small the 
galaxies could be suitable for the analysis. Fig. \ref{fig9_2} demonstrates the stellar disk scales and the 
central surface brightness (in arbitrary linear units) estimated from synthetic images. We created 21
images for a set of scale heights, and estimated resulting structural parameters from our 1-D and 3-D analysis. 
As \ref{fig9_2} suggests, reliability of both approaches is bad when the scale height is comparable with 
the pixel resolution in the images. The 1-D analysis overestimates the disk thickness by 15\%, given 
sufficient pixel resolution, which is due to the combination of non-perfect edge-on galactic inclination
and effects of the dust layer. The difference in the scales determined using these two approaches
can be seen in the real data in \S~4.1.

\begin{figure}
\epsscale{.80}
\plotone{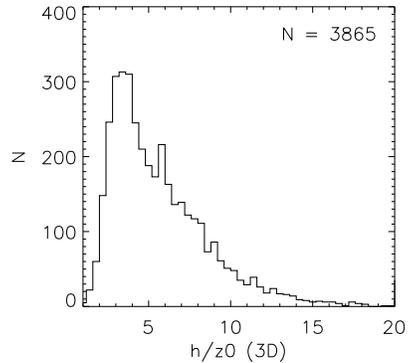}
\caption{The inverse r-band thickness of the stellar disks estimated in our 3-D modeling \label{fig11}}
\end{figure}

The distribution of the radial-to-vertical scale ratio is shown in Fig. \ref{fig11} for relatively large galaxies
(with $z_0 > $1.2 arcsec). The stellar disks look thinner from the 3-D analysis results in the comparison with
Fig. \ref{fig3}. The disk thickness estimated from different bands with the 3-D analysis looks similar.  The 
median inverse disk thickness is 5.7, 5.8, and 5.6 in the g, r, and i bands, respectively. 

\begin{figure}
\epsscale{.80}
\plotone{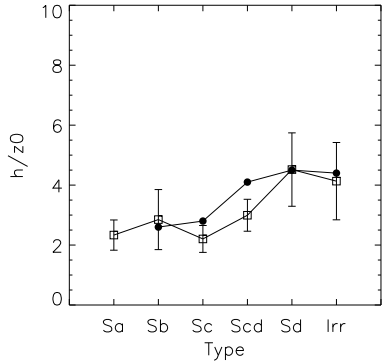}
\caption{The inverse stellar disk thickness for different morphological types in our sample based on 
our 1-D modeling is shown with the open squares with error bars. The filled circles show
the same results by \citet{degrijs98}.
\label{fig12a}}
\end{figure}

\begin{figure}
\epsscale{.80}
\plotone{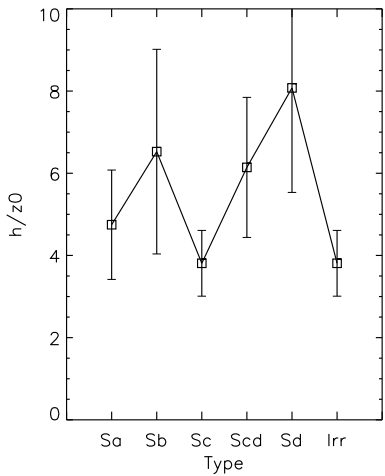}
\caption{ The same as Fig. \ref{fig12a} based on our 3-D analysis.
\label{fig12b}}
\end{figure}

Since morphological dependence of the stellar disk thickness has been reported from smaller sample studies \citep{degrijs98,schwarzkopf00},
we check how the average disk thickness is correlated with our preliminary morphological classification.
Figures \ref{fig12a} and \ref{fig12b} compare the median inverse disk thickness estimated for different morphological types.
Fig. \ref{fig12a} uses the results of our 1-D analysis, whereas Fig. \ref{fig12b} is based on 
the results from the 3-D analysis. 
It can be seen that Fig. \ref{fig12a} is in a very good agreement with figure 6 from \citet{degrijs98} 
(shown with filled circles) and close to the results
by \citet{schwarzkopf00}. Fig. \ref{fig12b}  reveals thinner stellar disks in the late-type spiral galaxies. It also suggests
that the galactic extinction contributes to the formation of the trend seen in Fig. \ref{fig12a}.

\begin{figure}
\epsscale{.80}
\plotone{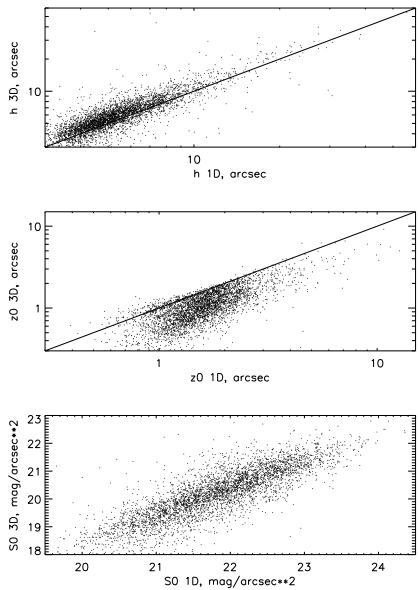}
\caption{Comparison between the structural parameters of the stellar disks estimated via 
the 3-D modeling and from the 1-D profile analysis. The panels show the radial disk scale length (top), 
scale height (middle) and the face-on surface brightness (bottom) estimated for the r-band images. \label{fig10}}
\end{figure}

The stellar disk structural parameters are estimated for the same galaxies by the 1-D and 3-D approaches 
independently from each other, so it is worth to compare  the resultant parameters. 
Fig. \ref{fig10} shows the comparison for the stellar disk scale length $h$, scale height $z_0$ and face-on central surface 
brightness $S_0$. The vertical scale height in the same images is smaller in the 3-D case, which suggests
that taking the dust into consideration improves the analysis. In combination with the correction of the
central surface brightness for the internal extinction, our 3-D approach produces significantly brighter stellar disks
in the comparison with the 1-D analysis. 

\begin{figure}
\epsscale{.80}
\plotone{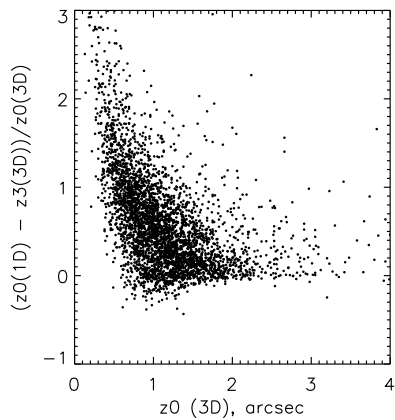}
\caption{The ratio of the stellar disk thickness estimated via 
the 3-D modeling and from the 1-D profile analysis for the r-band images.
While there is a nearly linear dependence for large values, there is a poor
agreement between the scales below 1 arcsec.
\label{fig10_2}}
\end{figure}

Both our 1-D and 3-D approaches to the modeling are affected by limited spatial resolution of SDSS images, which 
is severe for small galaxies. 
While deep optical and NIR observations of nearby galaxies allows for seeing very thin and low-contrast 
disk subsystems as a disk of blue stars reported by \citet{ultra_tinn_stellar_disks_n891}, most of our galaxies are observed with
rather limited resolution (the best SDSS seeing of 1 arcsec corresponds to approximately 1 kpc
at 15,000 km/s), which makes study of very thin subsystems impossible with our sample. Same reason 
prevents us from attempts of selecting the best functional form describing the vertical brightness profiles in thin disks
of different galaxies. Fig. \ref{fig10_2}  demonstrates the relationship between our 1-D and 3-D scale heights. 
It can be seen that although the trend in Fig. \ref{fig10} can be described as $z_{0,3D} = 0.58 +  0.94 \cdot z_{0,1D} $,
the vertical scales less than $\approx$ 1 arcsec deviate significantly from this linear relation. Both 1-D and 3-D ways of 
the scale estimation should be biased for the smallest galaxies in our sample because of the limited angular resolution.

The structural parameters of the stellar disks in the r-band determined with our 3-D modeling approach are shown in 
Table~6. 

\begin{figure}
\epsscale{.80}
\plotone{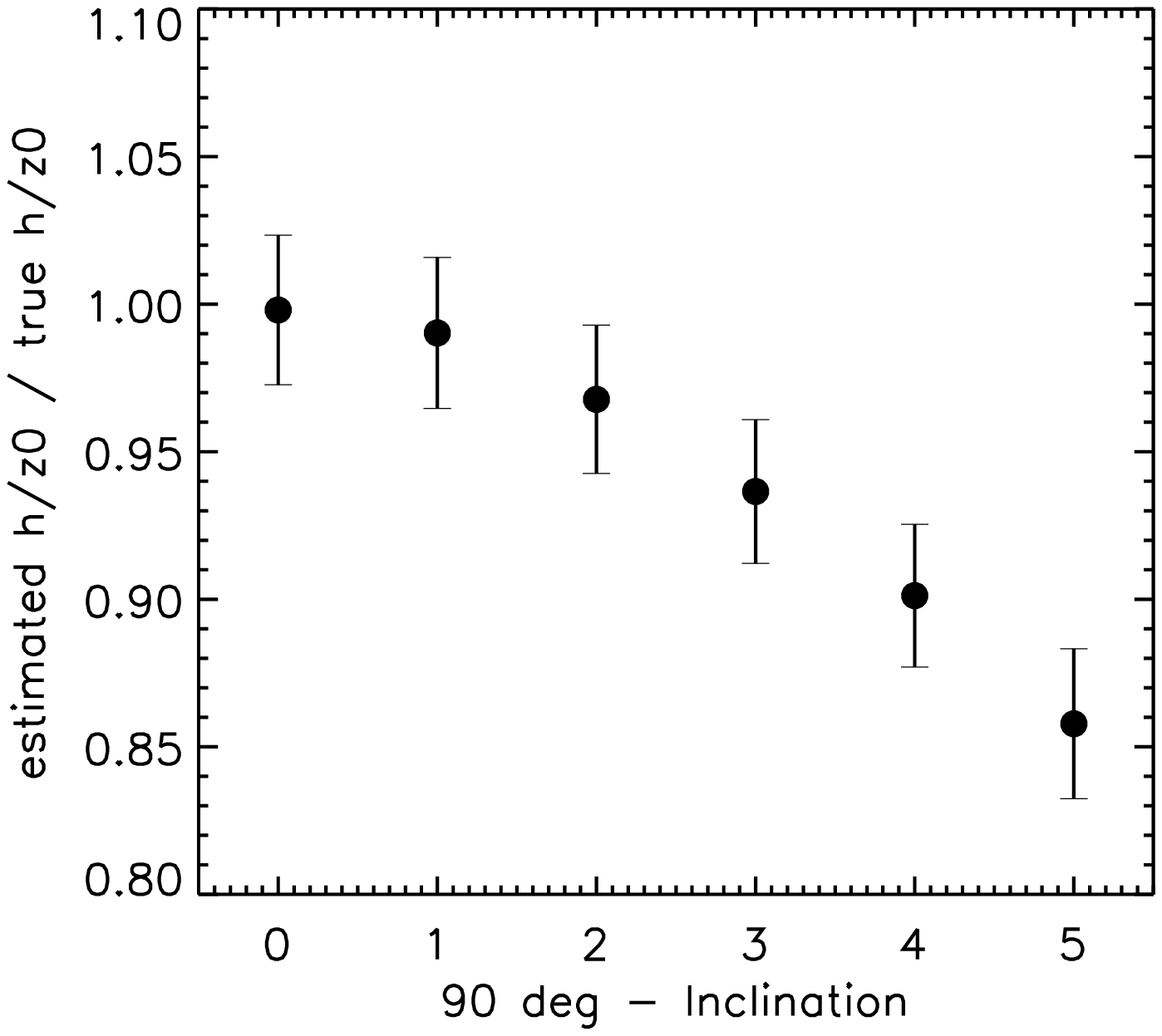}
\caption{The inverse stellar disk thickness $h/z_0$ estimated with our 1-D approach 
for a set of synthetic images with introduced non-zero inclination of the galactic
midplane. The original model $h/z_0$ was equal to 4, and there is no dust in the models.
The error bars shown in the picture are 10 times expanded and correspond to the 
models with central SNR = 50.
\label{fig18}}
\end{figure}

\textbf{Although the amplitude of uncertainties in $h/z_0$ ratio estimated with our 1-D approach 
is less than our typical observing errors, we can estimate the systematic addition to the 
original $h/z_0$ introduced by the non-perfect edge-on inclination of the galactic midplane to the
line of sight. We used the synthetic models developed in \S4 and estimated the disk thickness
$h/z_0$ with our 1-D approach with respect to the original one introduced to the artificial
models. We ran the synthetic images of edge-on galaxies with added noise (corresponding
to SNR = 50 at the center) through our 1-D analysis code. As Fig.~\ref{fig18} shows, deviation
of the inclination angle from the perfect 90 degrees introduces a few percent error in 
the disk thickness estimated with the 1-D approach.
} 

\subsection{Limitations to our estimations of the structural parameters}

We have to mention the limitations of our analysis of the galactic structure that originates from
the limited angular resolution, relatively low signal-to-noise of the data, as well as by simplifications
assumed for the analysis. 

Possibility of the multi-exponential stellar disk is a next step in complication of our 3-D modeling, together with introducing
disk warps and gradients of the disk scale height. Since this cannot be done for the  smallest galaxies in the sample, we 
will consider possibilities for a more complicated modeling in a next paper. 

Including the dust scattering into the modeling requires superior quality optical images, and the analysis remains very uncertain 
unless a multi-wavelength approach is developed \citep[e.g.][]{baes11,schechtman12}. Even in this case the realistic 
inclusion of the radiative transfer requires careful evaluation of each object individually. 
Consideration of more sophisticated shortcuts 
than plain neglecting the dust scattering and clumpy features in the dust layer in the modeling, which can be applied 
for our large sample, is a goal of our next work in this direction.  

A fundamental limitation to our attempt of separating the galactic structural components 
based on their different spatial scales comes from the fact that the thickness of the stellar disk disk in
low-massive galaxies is the same as that of the gas disks.  As it was noticed by \citet{dalcanton04},
disk edge-on galaxies with maximum of the rotation curve less than 120 km/s do not show 
regular dust lane and look mostly clumpy. Including this fact into our modeling requires kinematic 
information, or at least availability of the distances  for all galaxies in the catalog. We bring consideration
of this question out of scope of this paper and notice that our relatively small group 
morphologically classified as Irr, may mostly consist of such clumpy galaxies.

Our study does not attempt to provide precision modeling for all galactic components, neither select the best
individual model description to each object from a variety of available models. Instead it provides a ground for 
further studies of large samples of edge-on galaxies using both dedicated observations of limited
subsamples of certain groups of the galaxies and results from deeper observations that will 
come from large sky surveys. Incorporating the multi-wavelength information and of kinematic
data for a large fraction of the galaxies in our sample will improve the separation of the structural components.

\section{Conclusion}

Careful selection of candidate galaxies from SDSS images 
allows us to create the largest modern sample (5747 objects) of edge-on galaxies ready for further analysis. 
Our sample is complete for all galaxies with major axis larger than 30 arcsec. The distribution of the
the axial ratio shows that our sample size is reasonable, what also confirms its statistical completeness. 

We perform a 1-D radial and vertical profile analysis and infer the stellar disk's structural parameters. 
The results suggest that dust significantly biases the inferred parameters estimated from the optical 
band images.
We also perform a simplified 3-D 
modeling of all our galaxies taking into account the presence of dust. Comparison between the structural 
parameters shows that more constrained modeling is needed to eliminate effects of dust in the galaxies. 

The catalog can be used for statistical studies of the properties of the thick disks using stacked co-adding 
images. Our large sample makes possible studying scaling relations for galactic stellar disks and bulges.

\acknowledgments

We thank AAS Small Research grant program and grants RFBR-11-02-12247, RFBR-11-02-00471, 
RFBR-12-02-00685, and RFBR-14-02-00810 for partial support.
Funding for this project was partly provided through the Nova Southeastern University 
President's Faculty Research and Development Grant. We thank the 
anonymous referee whose multiple reviews significantly improved the paper. 

Funding for SDSS-III has been provided by the Alfred P. Sloan Foundation, the Participating Institutions, 
the National Science Foundation, and the U.S. Department of Energy Office of Science. 

SDSS-III is managed by the Astrophysical Research Consortium for the Participating Institutions of the SDSS-III 
Collaboration including the University of Arizona, the Brazilian Participation Group, Brookhaven National Laboratory, 
University of Cambridge, Carnegie Mellon University, University of Florida, the French Participation Group, the German 
Participation Group, Harvard University, the Instituto de Astrofisica de Canarias, the Michigan State/Notre Dame/JINA 
Participation Group, Johns Hopkins University, Lawrence Berkeley National Laboratory, Max Planck Institute for 
Astrophysics, Max Planck Institute for Extraterrestrial Physics, New Mexico State University, New York University, 
Ohio State University, Pennsylvania State University, University of Portsmouth, Princeton University, the Spanish 
Participation Group, University of Tokyo, University of Utah, Vanderbilt University, University of Virginia, 
University of Washington, and Yale University.

We acknowledge the usage of the HyperLeda database (http://leda.univ-lyon1.fr)

{}

%[2] Kautsch, S. J. 2009, AN, 330, 100 
%[3] Kautsch, S. J., et al. 2006, A&A, 445, 765 
%% Schechtman-Rook, Andrew; Bershady, Matthew A.; Wood, Kenneth	
%% 2012ApJ...746...70S

\clearpage

\begin{deluxetable}{lcrrrrrrrrrrrrr}
\tablewidth{0pt}
\tabletypesize{\scriptsize}
\rotate
\tablecaption{The Structural Parameters of True Edge-on Galaxies From The 1-D Analysis.}
\tablehead{
\colhead{Name} & 
\colhead{band} & 
\colhead{RA(J2000)} & 
\colhead{Dec(J2000)} & 
\colhead{PA} & 
\colhead{h} & 
\colhead{dh} & 
\colhead{z$_0$} & 
\colhead{dz$_0$} & 
\colhead{S$_0$} & 
\colhead{dS$_0$} & 
\colhead{grad(z$_0$)}& 
\colhead{mag} & 
\colhead{B/T} &
\colhead{Type}   
}
%\tabletypesize{\scriptsize}
%\rotate
\startdata
EON\_0.183\_7.090       & g &   0.182590 &   7.090147 &  157.78 &   7.31 &   0.61 &   1.21 &   0.19 & 23.75 &  0.26 &  -0.194 & 17.69 & 0.14 &  SB      \\
EON\_0.183\_7.090       & i &   0.182590 &   7.090147 &  157.78 &   5.56 &   1.14 &   1.30 &   0.20 & 22.82 &  0.50 &  -0.104 & 16.49 & 0.18 &  SB      \\
EON\_0.183\_7.090       & r &   0.182590 &   7.090147 &  157.78 &   7.61 &   2.56 &   1.31 &   0.18 & 23.51 &  1.04 &  -0.230 & 16.88 & 0.09 &  SB      \\
EON\_0.187\_33.757      & g &   0.186788 &  33.756809 &  156.87 &   2.55 &   0.17 &   0.91 &   0.09 & 21.80 &  0.15 &   0.012 & 18.23 & 0.08 &  SB      \\
EON\_0.187\_33.757      & i &   0.186788 &  33.756809 &  156.87 &   3.16 &   0.06 &   0.85 &   0.09 & 21.30 &  0.28 &  -0.048 & 16.90 & 0.15 &  SB        \\
EON\_0.187\_33.757      & r &   0.186788 &  33.756809 &  156.87 &   3.59 &   0.53 &   0.88 &   0.10 & 22.06 &  0.19 &   0.061 & 17.36 & 0.11 &  SB          \\
\multicolumn{15}{l}{...}\\
\multicolumn{15}{l}{Table is published in its entirety in the electronic edition.}\\
\enddata
%%\hbox{
\tablerefs{Parameters of the galaxies in the table: EGIS name, SDSS band, RA(J2000) in decimal degrees, Dec(J2000) in decimal degrees, 
position angle, scale length in arcsec and its uncertainty, scale height in arcsec and its uncertainty, face-on central surface brightness and its
uncertainty, vertical gradient of the scale height normalized by the scale ratio ( dz$_0$/dr * (h/z$_0$)), total uncorrected magnitude of the 
galaxy estimated by integration within the encompassing ellipse, average surface brightness of the galaxy within the encompassed ellipse,
the bulge-to-total ratio, mosphological type of galaxies, heliocentric
radial velocity in km/s (from LEDA; -1 is inserted if the value is unknown), and an alternative name. 
}
\end{deluxetable}

\clearpage

%%% Table 2.
\setcounter{table}{4} 
\begin{table} 
\begin{center}
\caption{The structural parameters derived from the 1-D profile analysis}
\begin{tabular}{lccc}
\tableline\tableline
 Parameter &  $g$ &  $r$ & $i$\\
\tableline
 h/z$_0$  & 3.57 $\pm$  1.61 & 3.31 $\pm$  1.31 & 3.17 $\pm$ 1.21 \\
 S$_0$    & 22.37 $\pm$ 0.81 & 21.56$\pm$ 0.81 & 21.10$\pm$0.84 \\
 B/T         & 0.24  $\pm$  0.16 &  0.29 $\pm$ 0.17 & 0.31  $\pm$ 0.17 \\
 h,arcsec$^*$      & 5.9  $\pm$  2.3 &  5.5 $\pm$ 2.0 & 5.3  $\pm$ 1.9 \\
 z$_0$,arcsec     & 1.8  $\pm$  0.6 &  1.8 $\pm$ 0.5 & 1.8  $\pm$ 0.5 \\
\tableline   
\end{tabular}
\vbox{
\setcounter{table}{4}
\caption{
Mean values and uncertainties of the reverse thickness h/z$_0$, central surface brightness,  and bulge-to-total 
ratio estimated for different SDSS bands. (*) Median value for the scale length and scale heights 
in arcsec are given with their uncertainty calculated as 1.48 * MAD (median absolute deviation).
}}
\end{center}
\end{table}

\clearpage

\clearpage

\begin{deluxetable}{lccc}
\tablewidth{0pt}
\tabletypesize{\scriptsize}
%%%\rotate
\tablecaption{The Structural Parameters of True Edge-on Galaxies in the r-band From Our 3-D Analysis.}
\tablehead{
\colhead{Name} & 
\colhead{h} & 
\colhead{z$_0$} & 
\colhead{S$_0$}
}
%\tabletypesize{\scriptsize}
%\rotate
\startdata
EON\_113.799\_20.000  &   8.86  &   1.04  & 21.34 \\
EON\_115.757\_45.121  &   6.53  &   2.09  & 19.77 \\
EON\_117.764\_50.255  &   3.97  &   1.13  & 19.22 \\
EON\_119.936\_45.366  &   6.77  &   1.54  & 19.46 \\
EON\_119.977\_47.413  &  20.66  &   0.05  & 20.79 \\
EON\_120.940\_15.240  &   4.92  &   0.03  & 18.84 \\
\multicolumn{4}{l}{...}\\
\multicolumn{4}{l}{Table is published in its entirety in the electronic edition.}\\
\enddata
%%\hbox{
\tablerefs{Parameters of the galaxies in the table: EGIS name (same as in Table~4),
scale length in arcsec, scale height in arcsec, and face-on central surface brightness in the r-band (mag~arcsec$^{-2}$). 
}
\end{deluxetable}

\clearpage

\end{document}